# Non-Inductive Current Start-Up Using Multi-Harmonic Electron Cyclotron Wave and Current Ramp-Up Through Combined Electron Cyclotron Wave and Ohmic Heating in EXL-50U Spherical Torus

Xinchen Jiang,[1,2,3] Yuejiang Shi,[2,3,*] Yueng-Kay Martin Peng,[2,3] Shaodong Song,[2,3] Wenjun Liu,[2,3] Xianming Song,[2,3] Xiang Gu,[2,3] Ji Qi ,[2,3] Dong Guo,[2,3] Debabrata Banerjee,[4] Lili Dong,[2,3] Zhenxing Wang,[2,3] Chunyan Li,[5] Junquan Lin,[6] Pingwei Zheng,[6] Haojie MA,[7] Huasheng Xie,[2,3] Jiaqi Dong,[8] Qingwei Yang,[8] Yunfeng Liang,[9] Baoshan Yuan,[2,3] Xianmei Zhang,[1,†] Minsheng Liu,[2,3] and EXL-50U team

[1]School of Physics, East China University of Science and Technology, Shanghai 200237, China

[2]Hebei Key Laboratory of Compact Fusion, Langfang 065001, China

[3]ENN Science and Technology Development Co., Ltd., Langfang 065001, China

[4]DISAT, Polytechnic University of Turin, Torino 10129, Italy

[5]School of Nuclear Science and Technology, University of South China, Hengyang 421001, China

[6]School of Resource Environment and Safety Engineering, University of South China, Hengyang 421001, China

[7]School of Liberal Arts and Sciences, North China Institute of Aerospace Engineering, Langfang 065000, People's Republic of China

[8]Southwestern Institute of Physics, Chengdu 610225, China

[9]Forschungszentrum Jülich GmbH, Institute of Fusion Energy and Nuclear Waste Management - Plasmaphysik, 52425 Jülich, Germany

The non-inductive current start-up by multi-harmonic electron cyclotron wave (ECW) has been systematically investigated in the EXL-50U spherical torus. Significant enhancements of the driven current with increasing number of resonance layers have been demonstrated by variation of the number of harmonic resonance layers of the ECW through adjustment of the magnetic field or plasma cross section. The critical role of multi-harmonic ECW in enhancing the driven current has been experimentally verified for the first time. To explain the related experimental observations, a physical mechanism involving multi-harmonic heating, multiple reflections, and multi-pass absorption—leading to the generation of high-energy electrons via X-mode wave or electron Bernstein wave (EBW)—has been proposed. The current drive capacity of the first harmonic extraordinary mode (X-mode) of the ECW has also been experimentally confirmed for the first time. After the application of Ohmic heating during the current ramp-up phase, the current drive efficiency of ECW is further enhanced. Leveraging the synergistic effect between ECW and Ohmic heating, EXL-50U achieved a plasma current of 1 MA, with the non-inductively driven current fraction reaching 70%.

The spherical torus (ST) device offers high performance within a compact design [1,2]. However, its low aspect ratio limits the space for the central solenoid (CS), restricting the available volt-seconds (V-s). A non-CS start-up method in STs was developed to reduce the loop voltage needed for plasma initiation. Electron cyclotron wave(ECW)-only start-up extends the ECW pre-ionization process, leading to [3]co-current electron orbit confinement, toroidal current generation, and magneto-hydro-dynamics equilibrium. Rapid current increases indicate the high efficiency of this process [4]. Future ST devices with superconducting coils will need to control the inductive loop voltage during breakdown, ramp-up, and flattop phases [5]. Several devices, including CDX-U [6], LATE [7], MAST [8], QUEST[9], and JT-60U [10], have successfully demonstrated non-CS start-up using ECW.

*Shi Yuejiang: yjshi@ipp.ac.cn

†Zhang Xianmei: zhangxm@ecust.edu.cn

Thousands of plasma pulses conducted on the EXL-50 device have demonstrated that the plasma current can stably reach 150 kA using ECW without the CS[11,12]. A preliminary hypothesis concerning the coupling between multi-harmonic resonance layers (1$^{st}$-5$^{th}$) and multiple reflections, resulting in multi-pass absorption, was proposed. Energetic electrons (ranging from several tens to hundreds of keV) play a unique and critical role in EXL-50, the metallic vacuum chamber walls, through multiple reflections and absorption effects under multi-harmonic resonances. It is suggested to use a low-field-side launched extraordinary mode (X-mode) when the electron cyclotron (EC) resonance is located at the fundamental resonance layer, providing more than two orders of magnitude higher electron cyclotron current drive (ECCD) efficiency than conventional ECCD in the sub-keV start-up regime [13]. Multi-harmonic electron cyclotron resonance heating and current drive (ECRH & CD) scenarios, including the fundamental and second harmonic layers, have also been proposed for ST-40 [14]. However, these hypotheses and theories have not been directly verified experimentally. By actively controlling the number of harmonic resonance layers on the EXL-50U, through adjusting the toroidal field ($B_T$) or the position of the outer limiter ($R_{out}$), we have directly confirmed the critical importance and necessity of multi-harmonic resonance layers and multiple reflections.

Based on these previous research, we submitted a model to explain experimental observations in EXL-50 and EXL-50U. As shown in Fig. 1(a), ray-tracing calculations show the propagation and absorption characteristics of the injected ECW, displaying the trajectories of the ordinary mode (O-mode in blue) and X-mode (red). The ECW frequency matches the EC frequency corresponding to the $B_T$. Under low-density start-up conditions, the X-mode can be efficiently absorbed. The O-mode, upon reflection on the vacuum chamber, is partially converted to the X-mode and absorbed [15]. Through the Doppler broadening mechanism, energetic electrons efficiently absorb ECW power across the broad yellow-shaded region and continue to accelerate as they pass through multiple harmonic resonance layers [12,16]. Multiple reflections of ECW inside the vessel allow the X-mode, along with the X-mode converted from the unabsorbed O-mode, can be transformed into electron Bernstein waves (EBWs) via the X-B process from the high-field side. These EBWs are then strongly absorbed at the upper hybrid resonance (UHR) layer, represented by the pink triangular shading [17]. According to this model, regardless of whether the initial injected ECW mode is X or O, after multiple reflections and multi-pass absorption, the wave will be absorbed in the form of either X-mode or X-B mode. When the wall absorption power is negligible, the mode of the initial wave does not influence the plasma current.

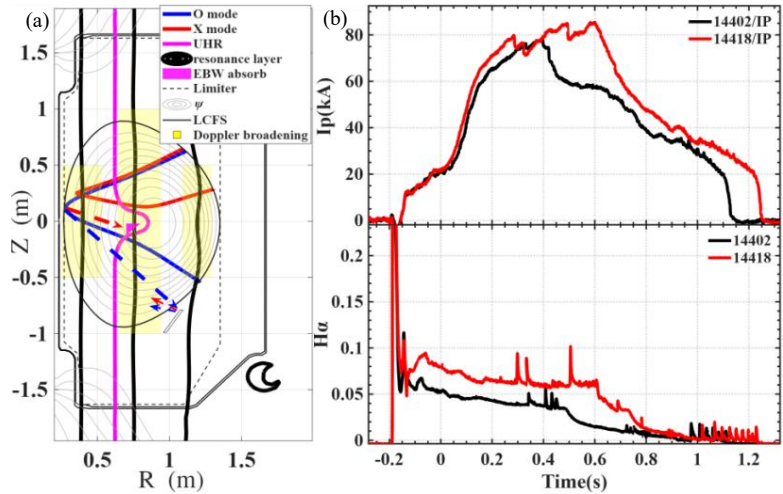

FIG 1. (a) Schematic of EC propagation and absorption in EXL-50U. (b) Pure O-mode (shot #14402) and pure X-mode (shot #14418) Non-CS start-up.

Experiments were performed on the EXL-50U [18,19] with a major radius of $R_0$ = 0.6 ~ 0.8 m and a minor radius of a = 0.45 m. The device was equipped with three 28 GHz EC systems (EC0, EC1 & EC2), as well as two 50 GHz systems (EC3 & EC4). Each system can deliver up to 200 kW in 2 s pulses. The $B_T$ was varied from 0.56 T to 1.2 T by adjusting the toroidal field coil current ($I_{TF}$) according to the device-specific calibration $B_T = 1.2 \times I_{TF} / 150$, where the denominator 150 corresponds to a maximum $B_T$ coil current. This variation effectively shifted the fundamental and higher-harmonic EC resonance layers across the plasma cross section. An adjustable outer limiter, positioned between $R_{out}$ = 1.000 m and 1.421 m, defined the plasma boundary and confinement region.

*Shi Yuejiang: yjshi@ipp.ac.cn

†Zhang Xianmei: zhangxm@ecust.edu.cn

As shown in Fig. 1(b), using 50 GHz EC4 injection through the lower launch, $R_{out}$ = 1.421 m, with identical gas fueling, EC power (200kW), and poloidal field (PF) current, pure O-mode (shot #14402) and pure X-mode (shot #14418) start-up experiments were conducted at $I_{TF}$ = 130 kA. In the presence of multiple resonance layers, both discharges reached plasma currents of about 80 kA. Analysis of the Hα emission indicates that the line integrated densities in these two cases are comparable at $n_e \sim 1 \times 10^{18}$ m$^{-2}$, and the evolution of the O- and X-mode ECW shows no significant difference. This consistently supports the physical picture of multi-pass absorption of the ECW by multiple reflections.

Under the 28 GHz EC0 injection conditions shown in Fig. 2(a) and 2(b), the position of the resonance layers was systematically varied by adjusting the $I_{TF}$ to 75, 110, and 130 kA, while the outer limiter remained fixed at $R_{out}$ = 1.200 m. When $I_{TF}$ was 130 kA, only the fundamental resonance layer stayed inside the vacuum vessel, resulting in the lowest plasma current, with a peak current of 25 kA. (Shot #14155). At $I_{TF}$ = 110 kA, the 28 GHz EC harmonic resonance layer shifted toward the central column, with two resonance layers present, and the current increased to 45 kA (Shot #14156). For $I_{TF}$ = 75 kA, three harmonic resonance layers coexisted, further increasing the current to 78 kA (Shot #14159). As the number of resonance layers increased, the plasma current showed a systematic rise.

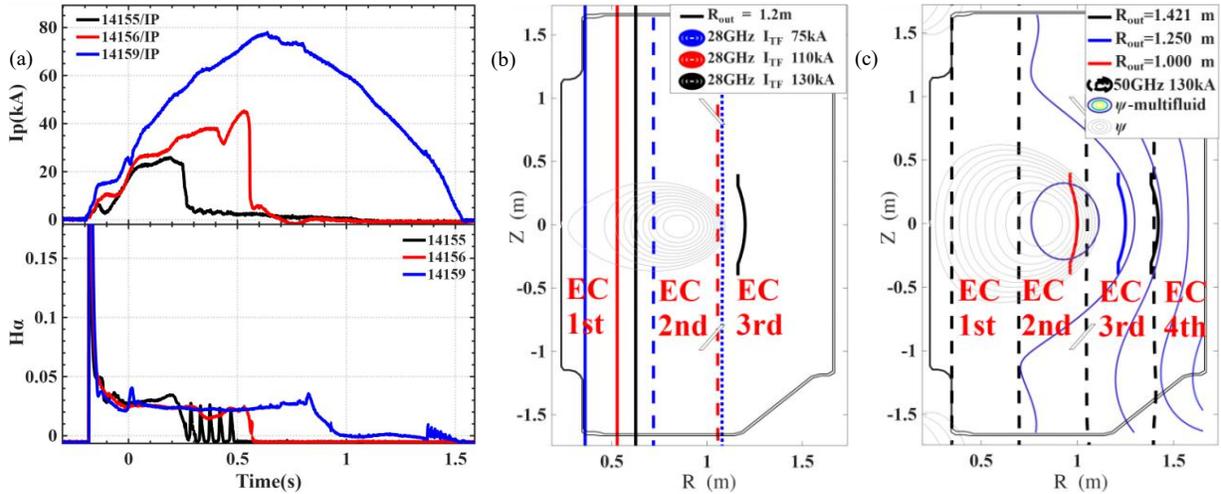

FIG 2. (a) Time evolution of Ip & Hα under EC0 injection. (b) Multi-harmonic resonance layers in $I_{TF}$ = 75,110,130 kA. (c) Multi-harmonic resonance layers in different $R_{out}$.

To further verify and expand these findings, the position of the outer limiter was adjusted (set to $R_{out}$ = 1.000 m, 1.250 m, and 1.421 m, respectively), and 50 GHz EC3 was used for start-up as shown in Fig. 2(c). The number of accessible harmonic resonance layers increases from the second to the fourth as the limiter position varies across the three locations.

*Shi Yuejiang: yjshi@ipp.ac.cn

†Zhang Xianmei: zhangxm@ecust.edu.cn

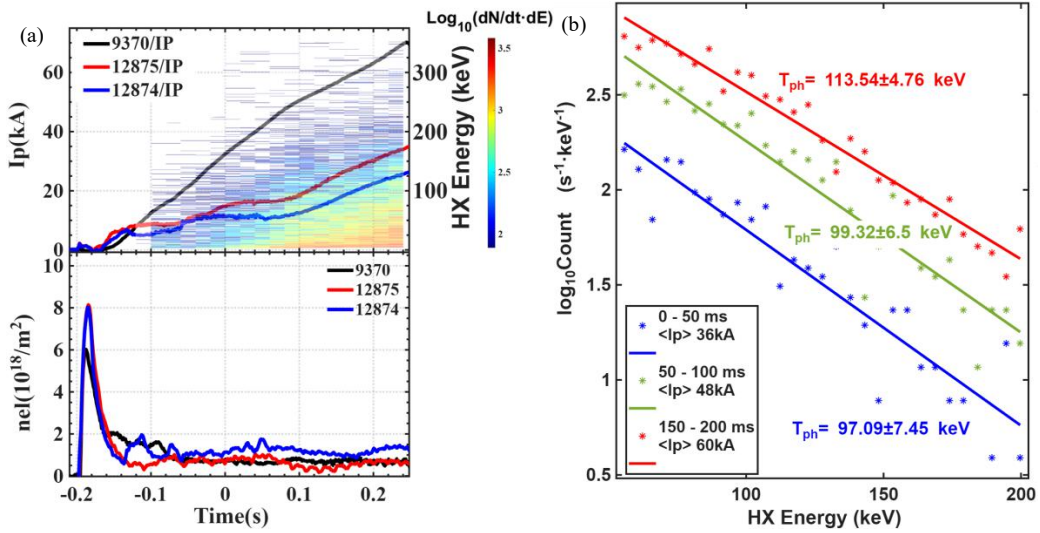

FIG.3. (a) Ip, $n_e$ and energy distribution in different $R_{out}$. (b) #9370 HXR spectra evolution.

Fig. 3(a) shows that plasma current reaches 26.9 kA when the limiter is at $R_{out}$ = 1.000 m (#12874), increases to 60.7 kA at $R_{out}$ = 1.250 m (#12875), and attains a record value of 96.3 kA at $R_{out}$ = 1.421 m (#12875). It also presents the energy distribution of energetic electrons for discharge #9370. By overlaying the time-energy 2D evolution map of hard X-ray (HXR) emission from #9370 onto the corresponding current waveform, it is clearly demonstrated that the EC-driven current is mainly carried by energetic electrons. Fig. 3(b) further shows HXR spectra extracted at three different time slices, revealing that as time progresses and the current increases, the photon temperature ($T_{ph}$, reflecting the characteristic energy of energetic electrons) stays nearly the same, while the HXR intensity rises significantly. This suggests that the current increase is primarily due to a rise in the number of energetic electrons.

The experimental results show that, while keeping other conditions constant, increasing the number of accessible harmonic resonance layers boosts the plasma current. The capacity of the fundamental X-mode to drive non-inductive current has also been experimentally clearly confirmed for the first time on the EXL-50U device. This finding provides a critical reference for the non-inductive current start-up in ITER-like devices, which has only the fundamental ECW layer. It is worth noting that, in contrast, the EXL-50 device has accessible 1st - 5th harmonic resonance layers. Due to the lack of passive stabilizing plates and the resulting expanded absorption region for energetic electrons, the EXL-50 achieves a higher plasma current during the non-CS start-up phase.

Non-CS start-up in which the key ingredients are the presence of multiple accessible harmonic resonance layers and the ability of energetic electrons to traverse sufficient spatial regions to absorb wave power and form current in Fig. 4(a). Multi-harmonic absorption enables efficient deposition of ECW energy, thereby increasing plasma parameters and producing a significant population of several tens to several hundreds of keV electrons whose perpendicular and parallel velocities are comparably enhanced and which carry the majority of the plasma current.

ECW has already driven a portion of the plasma current before the CS engages. Subsequently, the toroidal electric field provided by the CS, along with the NI contribution from ECCD, further increases the electron $v_{//}$, strengthening the overall current drive. A positive feedback loop between improved confinement and current generation is thus established, ultimately enabling current ramp-up. Through flux consumption synergy, the system can effectively manage the ramp-up rate and flat-top feedback.

*Shi Yuejiang: yjshi@ipp.ac.cn

†Zhang Xianmei: zhangxm@ecust.edu.cn

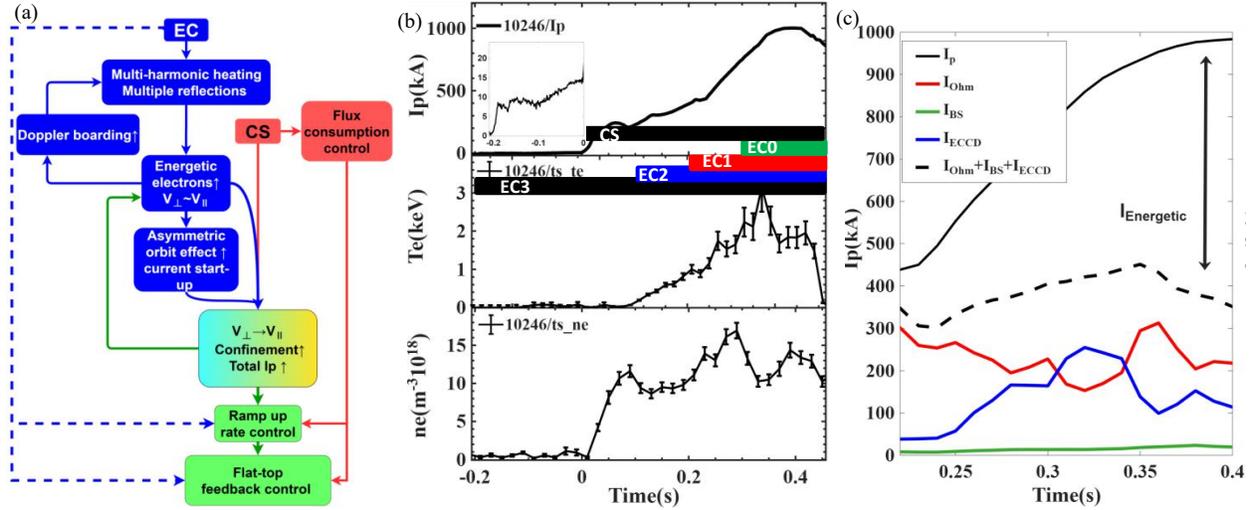

FIG.4. (a) Non-CS start-up by multi-harmonic electron cyclotron wave heating and current ramp-up flow chart. (b) Time evolution of 1 MA EXL-50U discharge. (c) Plasma current components: total plasma current (Ip), CS-induction current (Iohm), bootstrap current (IBS), and ECCD.

In the current design of the EXL-50U, a 200 ms EC discharge is initially applied, followed by a ramp-up of EC+ CS. The fastest non-CS ramp-up speed achieved so far is 97 kA in 500 ms, while the current ramp-up driven solely by EC waves is relatively slow, and the $B_T$ sustainment time in standard conductor devices is limited. As shown in Fig. 4(b), before CS operation, EC3 alone drives a current of 15 kA. Once the CS is engaged, the current ramps up at a rate of 3.2 MA/s. During the current increase to the flat-top phase, shown in Fig. 4(c), the CS contributes about 25% of the total current, while conventional EC (800 kW) based on thermal electrons contributes around 12%; however, the majority (~60%) of the plasma current is carried by energetic electrons, ultimately enabling the achievement of 1 MA. In the 1 MA discharge, the plasma electron temperature reaches 3 keV, and the density reaches $n_e = 1 \times 10^{19}$ m$^{-3}$.

Historically, plasma current initiation and ramp-up in tokamak have mainly depended on inductive current drive via the CS. The EXL-50U experiment confirms a promising alternative approach, showing that ECW can achieve complete non-CS current start-up and play a major role during current ramp-up. Although the CS's role thus shifts, it remains a critical part for improving current drive efficiency, controlling the ramp-up rate, and providing current feedback during the flat-top phase. The EXL-50U results offer a promising path for current drive in future spherical and conventional tokamak reactors.

*Shi Yuejiang: yjshi@ipp.ac.cn

†Zhang Xianmei: zhangxm@ecust.edu.cn

*Shi Yuejiang: yjshi@ipp.ac.cn

†Zhang Xianmei: zhangxm@ecust.edu.cn